\documentclass[12pt]{iopart}
\usepackage{epsfig}
\begin{document} 
%\draft

\title{DNA condensation and redissolution: Interaction between 
overcharged DNA molecules. } 

\author{Elshad Allahyarov \dag \ddag \S\ Gerhard Gompper\dag\ and
Hartmut L\"owen \ddag 
\footnote[1]{To
whom correspondence should be addressed (allahyar@thphy.uni-duesseldorf.de)}}

\address{ \dag\ Institut\, f\"{u}r\, Theoretische \,Physik
  II,\,Heinrich-Heine-Universit\"{a}t\,
 D\"{u}sseldorf,\,\mbox{D-40225}\,D\"{u}sseldorf, \,Germany}
\address{\ddag\ Institut\, f\"ur\, Festk\"orperforschung, Forschungszentrum 
J\"ulich, \,\mbox{D-52425} \, J\"ulich, Germany }
\address{\S\ Institute for High Temperatures, Russian Academy of
Sciences, Izhorskaya 13/19, \mbox{117419} Moscow, Russia}

\begin{abstract}
The effective DNA-DNA interaction force is  calculated by computer 
simulations with explicit tetravalent 
 counterions and monovalent salt. For overcharged DNA molecules, 
the interaction force shows a double-minimum structure. The positions
and depths of these minima are regulated by the counterion density in
the bulk. Using two-dimensional lattice sum and free energy perturbation
theories, the  coexisting phases for DNA bundles are calculated.  
A DNA-condensation and  
 redissolution transition and a stable mesocrystal with an intermediate lattice
 constant for high counterion concentration are obtained. 
   
\end{abstract}
\pacs{PACS: 87.15.Kg, 61.20Ja, 82.70.Dd, 87.10+e}

\maketitle

\section{Introduction}
Long DNA is naturally found in a dense form in most biological
systems due to the presence of compacting agents. In vitro the most
used agents are polyamines such as trivalent spermidine (Spd) and
tetravalent spermine
(Spe) \cite{cohen1998book}. These agents play a key role in
maintaining cellular DNA in a compact state
 \cite{bloomfield1997,saminathan},  modulate ion channel activities
 of cells, are essential for
normal cell growth \cite{cason2003spermine} and  can effectively be applied
in gene delivery and in the field of genetic therapy. 
 Under physiologic ionic and pH
conditions, the polyamines are positively charged and hence DNA is
their prime target of interaction. The molecular mechanism
of polyamine function in DNA condensation is presumed to involve
neutralization of the negatively charged DNA backbone by the
positively charged amino groups of Spd and Spe \cite{wilson1}.
Experimental results  and counterion condensation theories indicate 
that non-specific  interactions are
predominantly electrostatic between
polyamines and DNA phosphates
\cite{deng2000polyamine,braunlin1986biopolymers,olvera1995,raspaud1999}.
Thus, the electrostatic shielding of phosphates
facilitates close helix-helix surface contacts and, ultimately,
DNA condensation through the correlation attraction
\cite{Levine2002attarctingrods}, the attraction between strongly
correlated counterion layers on the adjacent DNA surfaces.

 In the last decade, different experiments
 \cite{saminathan,pelta1996,longtriplex1999saminathan,solis2001reentrance,solis2002,vries2001recondensation,murayama2003reentrance}
 have shown evidence of 
DNA redissolution (i.e. DNA unbinding)  with increasing concentration
 of polyamines and several of their
structural analogues. 
First, addition of a certain amount of multivalent salt causes the
collapse of single DNA or the bundling in multi-columnar
structures. Upon addition of more salt, the polyelectrolyte redissolves
and DNA assumes again an unbundled conformation. 
There are three experimentally well established features of
the redissolution phenomena: i) a  linear relationship is found between
the threshold concentration of multivalent ions $C_c$, when the onset of DNA
condensation takes place, and the initial DNA concentration $c_{DNA}$
 (the DNA concentration in solution free of multivalent ions);
 ii) the decondensation 
threshold $C_d$ of multivalent ions, when the condensed DNA returns back to 
solution, is almost independent of monovalent salt
concentration $c_s$; iii) between the threshold values  $C_c$ and $C_d$ the
 DNA fragments show two 
coexisting liquid crystalline  phases.  Concerning the first item, the onset of DNA
 condensation is usually explained by the correlation attraction between almost neutral
structures. Thus, the precipitation induced by trivalent or tetravalent
ions is not a consequence of the intrinsic structure and flexibility
of DNA, but is a common feature of a  polyelectrolyte solution.
The threshold value $C_c$  grows with increasing  monovalent salt concentration $c_s$ 
 \cite{bloomfield1997,nguen_reentrant,bloomfield1996cosb,burak2003}. 
A mono-molecular DNA
collapse into a neutral toroidal structure occurs in highly dilute
solutions of long DNA chains \cite{saminathan,raspaud1999}, whereas
a multi-molecular aggregation is generally observed in more concentrated DNA
solutions, regardless of the DNA length \cite{arscott1990dnacollapse,Solis2000}.
The second item, the DNA redissolution at $C_d$, is currently under
intensive debate in the colloidal community
with different, and sometimes conflicting, explanations
 \cite{longtriplex1999saminathan,solis2001reentrance,murayama2003reentrance,nguen_reentrant,trubetskoy2003reentrance}. 
For instance, 
in Ref.~\cite{longtriplex1999saminathan} it is argued that after precipitation the
increased binding of polyamines will make the DNA hydrophilic enough to dissolve in
water. Other experiments \cite{trubetskoy2003reentrance} show that the
DNA is still in a condensed state when polyions are added beyond
the threshold concentration $C_d$, but it is more finely dispersed
in the solution. In Ref.\cite{nguen_reentrant} the
reentrance is explained by resorting to DNA overcharging phenomena, which take
place roughly in the middle of the condensation plateau.  
DNA is claimed to experience 
negative electrophoresis and move opposite to the
conventional direction near the reentrance transition.
 However, experiments of
Raspaud et al \cite{raspaud1999} do not confirm this 
claim. 
In Ref.\cite{solis2001reentrance} it was suggested that the
redissolution is very sensitive to the relationship between the
condensation of multivalent counterions on the polyelectrolyte and
multivalent counterion--monovalent coion association (Bjerrum pairs). Thus, if
the chemical potential of the multivalent counterions is low, they more
likely adsorb on the DNA and overcharge it. On the other hand, if the 
chemical potential is high, 
the counterion-coion associations are  more likely to appear in solution. A resulting 
condensation of Bjerrum pairs  creates consecutive layers of oppositely charged ions
 around the DNA molecule \cite{solis2002,tanaka2001chargeinversion}.

While the first two above mentioned  items have been studied in
considerable detail,
much less attention has been paid to the third item, namely the
coexistence of two different liquid-like structures in DNA
condensates.
In a series of experiments, Livolant and colleagues demonstrated that
the spermidine and spermine ions are capable of provoking several liquid
crystalline forms of fragmented DNA \cite{pelta1996}. Similar
results were published recently by Saminathan et al in Ref.~\cite{saminathan}.

In this paper we investigate the condensation
and redissolution of DNA on a molecular level by using
computer simulations of the primitive-model electrolyte 
with explicit tetravalent counterions
and monovalent salt ions.
We trace back the condensation and redissolution to the 
distance-dependent effective potential $U(R)$ between two parallel DNA molecules,
where $R$ is the radial distance between their two centers.
Using two-dimensional liquid-state theory for the fluid and lattice sums for
the solid phases, we calculate the phase diagram for columnar DNA assemblies.
A previous account of the results was already published elsewhere
\cite{EPLpaper}.

 The remainder of this paper is organized as follows. In Section II we
 describe our model system and give the definition of DNA-DNA interaction forces. 
 We calculate the interaction forces for different counterion and salt 
concentrations in  Section III.
 Section IV is devoted to
 the free energy perturbation theory for
 defining the liquid-liquid coexistence densities.
We conclude in Section V.

\section{The model}
We consider a B-DNA molecule which is a double helix with a
 pitch length $P$=34\AA.  There
 are $N_p$=20 phosphate charges per one 
 helical turn which makes one elementary charge per each
1.7\AA \, of axial rise. The geometrical shape and charge distribution
 of DNA is modeled through the accurate Montoro-Abascal model (MAM)
 \cite{montoro1998,oursecondDNApaper}. Its cross section is
 illustrated in Fig.~\ref{figure_1}.

A pair of DNA molecules are placed 
along the {\it xy} diagonal of the cubic simulation
box of volume $V=L^3$. The size of the simulation box
$L=102$\AA \, corresponds to three full turns of DNA.
The box also consists of $N_Q$ multivalent ions with valency $q_Q$, $N_- = N_{s-} + q_Q N_Q$
monovalent coions and $ N_+ = N_{s+} + N_c$ monovalent counterions. Here 
$N_{s+}=N_{s-}=N_s$ is the number of added salt ion pairs, 
$N_c$ is fixed by the 
DNA phosphate charge due to the constraint
of global charge neutrality ($N_c=120$ in our simulations). 
All ions are modelled as charged hard spheres. 
Tetravalent counterions are assumed to represent
spermine ions. Though the latter is a
polyion in {\it vivo}, a spherical approximation can be used,  since
the bibliographical data support the idea that it is the charge of
counterion, rather than its structural specificities, which is important in
DNA condensation and redissolution processes \cite{raspaud1999}. 
 Beside of this, the fluidity of
the ordered DNA phase also suggests that spermine binds 
like an ordinary ion to the  DNA surface, rather than forming 
 inter-strand cross-links 
to neighboring DNA molecules \cite{pelta1996,longtriplex1999saminathan}.
Assuming that the ions of the same valency are indistinguishable, the actual
number of mobile ion species in the simulation box is reduced from five (which
are multivalent ions and their 
coions, positive and negative monovalent salt ions and  monovalent
 counterions that neutralize
the DNA phosphate charge) to three:
 multivalent counterions (charge $q_Q=4$ and diameter $d_Q$) and positive and
 negative small ions (charge $q_c=\pm 1$ and diameter $d_c$).
Periodic boundary conditions in all three
directions are applied to reduce finite size effects.
 The DNA replicas in the {\it z} direction produce infinitely long
molecules. The whole system is held at room
temperature $T=298K$ and the water is modelled as a continuous
dielectric medium with dielectric constant $\epsilon=80$.
A typical snapshot of the simulation is shown in
Figure~\ref{figure_2}.

The interaction potentials between
the five sort of particles (three of them are the mobile 
ions in solution mentioned above, and two 
of them, one charged and the other neutral, belong to the DNA
molecule, see Figure~\ref{figure_1}) are a combination of hard core and Coulomb potentials: 
\begin{equation}
V_{ij} (r) =\cases {\infty &for $ r \leq (d_i+d_j)/2 $\cr
   {{q_i q_j e^2} \over {\epsilon r}} &for $ r > (d_i+d_j)/2 $, \cr}
\label{1cLMH}
\end{equation}
Here $r$ is the  inter-particle distance, $i,j$= $Q$
for multivalent counterions,  $c$ for small ions, $p$ for phosphate charges 
 and $n$ for neutral spheres in the MAM (with $q_n$=0).
There is also an excluded volume potential $V^{0}_{i}$ between the
DNA hard cylinder and the free ions $i,j= Q,c$.

The basic quantity of interest is the effective
 force  per  helical turn \cite{ourfirstDNApaper} 
\begin{equation}
{\vec F} = {\vec F_1} + {\vec F_2} + {\vec F_3}.
\label{eq_1}
\end{equation}  
between  two DNA molecules. Here ${\vec F_1}$ is 
the direct Coulomb repulsion per helical turn  of one DNA molecule as
 exerted from the  
phosphate groups of the other DNA. It does not
 depend on salt density and its calculation is straightforward
 \cite{ourthirdDNApaper}.  
The second term  ${\vec F}_2$ in Eq.~(\ref{eq_1}) corresponds to Coulomb
interactions between the phosphate charges at positions
 ${\vec r}_k^{~p}$ ($k=1,..N_p$) and the mobile ions at positions ${\vec
r}_l^{~i}$ ($l=1,..N_i$, $i=c,Q$),
\begin{equation}
{\vec F}_2= - \frac{1}{3}{\sum_{k=1}^{3N_p}} \left( \langle \sum_{i=c,Q} \sum_{l=1}^{N_i}
 {\vec \nabla}_{{\vec r}_k^{~p}}  V_{pi}( \mid {\vec r}_k^{~p} - {\vec
 r}_l^{~i} \mid ) \rangle \right).
\end{equation}  
Here $<...>$ denotes  canonical
average over small ion  configurations.
 The third term ${\vec F}_3$ in Eq.~(\ref{eq_1}) arises from the entropic contribution of
small ions due to their moment transform to the DNA surface ${\cal S}$,
\begin{equation}
{\vec F}_3=-\frac{1}{3}k_BT \int_{{\cal S}} d{\vec f} \ \
\left(\sum_{j=c,Q} \langle \rho_j({\vec r}) \rangle  \right ).
\end{equation}  
Here ${\vec f}$ is a surface normal vector pointing outwards the DNA
core and $\rho_j$ ($j=c,Q$) is the inhomogeneous microion concentration. 
The canonical averages in ${\vec F}_2$ and ${\vec F}_3$ are carried
over all configurations of the mobile ions. 

We have performed extensive grand-canonical molecular dynamics (GCMD)
 simulations, similar to those of 
 Ref.~\cite{ourthirdDNApaper}, for a
 range of different tetravalent counterion  and monovalent salt concentrations.
Each simulation is referred by its nominal
tetravalent counterion concentration $C$ (salt ion concentration
 $c_s$) defined as a ratio between the total number of ions
$N_Q$ ($N_s$) in the cell without the DNA molecules and the system
 volume $V$, $C=N_Q/V$ ($c_s=N_s/V$). Additional simulations have been
 carried for these bulk phases in order to calculate the
 chemical potential $\mu$ of solution. Then 
in the simulations with DNA molecules  the number of ions
in the simulation cell was automatically
adjusted to the specified value of chemical potential $\mu$.  
The ion diameter was chosen to be $d_Q$=8\AA \, for tetravalent
counterions and $d_c$=4\AA \, for other free ions in the system. This
parameter defines the closest approach of the ion to the DNA surface
and has a strong impact on the polyion electrostatics. The time step
$\triangle{t}$ of 
the simulation corresponds to an average ion displacement of $0.03$\AA
 \, per time
step such that the reflection of
counterions following the  collision with the combined surface of DNA
is calculated with high  precision. 
About $5 \times 10^4$ MD steps are required on average to reach 
equilibrium. The time evolution is then followed  for $5\times
 10^4-5\times 10^6$ steps to gather statistics to calculate 
 canonical averages of the interaction forces.

\section{Results for the interaction forces}
The case of a single DNA molecule  in the presence of spermine ions has been
considered in Ref.~\cite{EPLpaper}. It was shown there (see
Figure 2 in Ref.~\cite{EPLpaper}) that the ionic cloud may not only
compensate the polyion charge but even exceed it, resulting in
an opposite sign of the electrostatic potential at some distances from
the DNA surface. 
The adsorption of Bjerrum pairs \cite{tanaka2001chargeinversion} 
 onto the DNA surface at high tetravalent counterion concentration creates
 consecutive layers of
charges of different sign around the DNA molecule. 
The onset of a multilayer structure occurs at  $C=1.8$mM. 
Addition of monovalent salt shifts this threshold concentration to lower
values of $C$, in accordance with experimental observations and 
two-component Manning condensation theory \cite{manning1978salt}.  
For multivalent counterion
concentrations exceeding $C=1.8$mM, the total charge in the DNA grooves remains
constant and only the  total ionic charge adsorbed on the strands
contributes to the overcharging, similar to our earlier findings
\cite{oursecondDNApaper,EPLpaper}.
Beside of this, there is a competition between the
multivalent  and monovalent counterions in
binding to the DNA surface as $C$ increases. 
 The multivalent ions tend to replace the monovalent
counterions on the DNA surface. Thus, at higher $C$
the main contribution to the formation of charged layers around DNA molecules comes
from Bjerrum association between big counterions and small coions. 
Such charged layers give rise to an attraction between two
parallel DNA molecules, as shown in
Figure~\ref{force_charge}. A decrease
 of the  big counterion charge leads to the break-up of 
the Bjerrum counterion-coion pairs  and  thus destroys the layer
formation around the DNA molecule.
 This ultimately results in the loss of intermolecular attraction,
 see simulation data for $q_Q=$1,2 and 3 in
 Figure~\ref{force_charge}. 

It should be mentioned that in addition to the intermolecular (or
 axis-to-axis) distance
 $R$, there are angular variables which define the mutual
 configuration of two parallel DNA molecules corresponding to 
the orientation of their grooves and strands
\cite{ourfirstDNApaper}. At short
surface-to-surface distances between the two DNA molecules up to 5\AA \, 
 there are strong contributions from particular DNA-DNA
 configurations \cite{ourthirdDNApaper}. 
 For larger separation distance, $R>25$\AA \, we find no detectable dependence of the
interaction forces on the mutual orientations of DNA molecules.
On the other hand, there is
 experimental evidence that in DNA condensates two
 neighboring molecules never approach each other closer than 5\AA. This
 apparently means that at such small distances a strong
 repulsion between DNA solvation shells exists.
Thus in all figures hereafter
we show orientationally averaged interaction forces
 starting from the distance $R=24$\AA. 
The interaction forces are scaled per 
DNA pitch, i.e. per 10 DNA base pairs. 
%Thus, for interaction forces at intermolecular distances below $R=24$\AA
% \, we adopt a $1/R^{12}$-like repulsion.

The electrostatic $F_2$ and entropic $F_3$ components
  of total interaction force  $F$
 for tetravalent counterions  corresponding to DNA
overcharging are separately plotted in Figure~\ref{figure_4}.  
It can be seen that the electrostatic force shows oscillations around
  zero, which are reminiscent of the  multilayered
structure of charges around a single DNA molecule \cite{EPLpaper}.
The deep attractive minimum of the total force $F$ has an 
entropic origin, whereas the second minimum at intermolecular distance
R=41\AA \,  has a purely electrostatic origin.

In Figure~\ref{force_salt} we plot the DNA-DNA interaction force for
both the undercharged and overcharged cases at different salt
concentrations. It is evident that in dense salt solutions the
attractive minimum becomes weaker. For overcharged DNA,
$C=65$mM, the position of the second minimum shifts towards the DNA
surface. Thus, whereas the DNA overcharging does not strongly depend
on the added
salt concentration $c_s$ \cite{EPLpaper}, the effective forces do.

The  positions of the
minimum and the maximum of the force shift towards the DNA surface
also for higher spermine 
concentrations.
 This trend is shown in
Figure~\ref{figure_5a}, where the DNA-DNA interaction force is plotted
for different spermine and salt concentrations at 
fixed  distance $R=38$\AA. For  low $C$,
 which in Figure~\ref{figure_5a} corresponds to 
the area to the left of point A, the DNA-DNA interaction is overall
repulsive.  
Between the points $A$ and $B$ a first minimum develops in $F$.  As the spermine
concentration increases further, the minimum shifts toward the DNA surface
and the force has a repulsive tail. This tail indicates the appearance
of second maximum, which obviously is followed by second minimum.

The dependence of the DNA-DNA interaction force on $C$ for two fixed
intermolecular distances are shown in Figure~\ref{force_min_max}.   
Five different arrows in this picture point to different values of $C$
which characterize the number of attractive minima in the interaction
force $F$. For small $C=0.01$mM, indicated by the arrow next to {\it a} in 
Figure~\ref{force_min_max}, the interaction force has no minimum and
thus is totally repulsive. For spermine concentration $C=0.025$mM,
corresponding to the arrow next to {\it b}, the force has a single minimum.
For intermediate $C=1.7$mM and $C=56$mM, see the arrows next to  {\it c} and {\it
d} respectively, there are two minima in the
interaction force (note that a positive force at $R=$ 38\AA
\, for $C$ corresponding to arrow {\it d} indicates 
the occurrence of second maximum, which obviously if followed by 
second minimum at larger $R$). And finally, at even higher $C=400$mM, the first minimum
has disappeared, however the second minimum is retained.

A full set of distance resolved DNA-DNA interaction force curves for 
different tetravalent counterion densities $C$ are presented in
Figure~\ref{interctionforce}. It can be seen that even a small trace
of spermine ions -- well below the overcharging threshold -- 
induces an attraction between the DNA molecules,
except at very close
distances, see curve for $C$=0.1mM.
As the DNA molecules get more overcharged, which
corresponds to high spermine concentrations, the  main
minimum narrows and becomes more shallow. At the same time, the width
and the height of 
the maximum increases. We note that the attractive
minimum for undercharged DNA pair has mainly
a pure electrostatic origin and arises due to charge correlations
in the electrolyte.
 However,
for overcharged DNA the main contribution to the force at 
this minimum is due to the spermine crowding near the DNA surface. The second maximum
originates both from electrostatic and entropic 
forces. Finally, the second minimum emerges from  a pure electrostatic 
effect.

In the following we
calculate the total effective pair interaction potential $U(R)$ per
unit length for a given bulk salt concentration
 $c_s$ and different Spe concentrations $C$. The quantity $U(R)$
 is obtained by integrating the distance-resolved interaction force
 averaged over all microion configurations \cite{allah}. 
Results are shown in Figure~\ref{figure_5}. The oscillations in the
force imply that the interaction potential also oscillates.
With increasing $C$, the first minimum of  $U(R)$ is getting deeper and
is achieving a maximal 
depth at the overcharging concentration $C \approx 1.8$mM. 
A further increase of $C$ again reduces
the depth of this minimum. The position of the minimum, on
the other hand,
hardly depends on $C$. 
The Spe-layering around the pair
of DNA molecules induces a second minimum at larger
separations as revealed in the enlarging inset of Figure~\ref{figure_5}.
This minimum is of electrostatic origin and occurs for $C \ge$65mM.
Again the depth of the second minimum first increases and then
decreases with increasing $C$. 
At intermediate Spe concentrations,
we are thus confronted with a double minimum potential which is induced
by layering. It is worth to mention that positions of the second minimum
can be related to intermolecular distances between the DNA molecules in 
cholesteric phase induced by  polyamines
\cite{durand1992cholesteric,saminathan}.
  Direct measurements \cite{strey1999chiral} and   theoretical
investigations \cite{KL2000cholesteric} of intermolecular forces
demonstrated that the energetics of this cholesteric phase is
determined primarily by electrostatic interactions. 
Since  the  potentials in
Figure~\ref{figure_5} are scaled for one DNA pitch length,
the interaction strength corresponding to the minimum of curve
(2) for
very long DNA molecules is sufficiently large to induce condensation. 
This implies  that  DNA aggregation can take
place well {\it below} overcharging  Spe concentrations.

\section{Phase diagram for double-minimum potential.}
The characteristic double-minimum structure of the interaction potential
$U(R)$ will give rise to an unusual phase behavior. We have calculated the
 phase diagram of a columnar DNA assembly on the basis of our
simulated effective pair interactions. To do so, we assume
that the DNA molecules are parallel along a certain length $\ell$.
This length is an additional parameter 
which we fix to be $\ell =20 \times P$. We comment on the dependence
of the phase diagram on $\ell$ in Section V. 
The assembly of parallel DNA can be considered as a two-dimensional
many-body system interacting via a potential $\ell \times U(R)$ and being
characterized by a DNA particle number density $\rho$. 
We calculated the free energies of the fluid
and solid phases by using different techniques outlined below and perform the
traditional Maxwell double tangent construction to identify the coexisting densities.

The free energy of dilute {\it fluid} phase is approximated by the 
two-dimensional perturbation theory \cite{wca1979} via splitting the total
potential into a repulsive and attractive parts, $U(R)= U_r(R) +
U_a(R)$. 
 The repulsive potential $U_r(R)$, identical to $U(R)$ but
 truncated and shifted towards zero 
 at the first minimum at $R=R_{min}$,  is  mapped
onto that of effective hard disks of diameter $\sigma_{eff}$ \cite{evans_leshouse}
\begin{equation}
\sigma_{eff} = \sigma +
\int_{\sigma}^{R_{min}}{ \left[1-\exp{\left(-\frac{U_r(R)}{k_BT}\right) } \right] }dR.
\end{equation}
Here the cross-section diameter for the DNA molecule is $\sigma=20$\AA.  
The total Helmholtz free energy involves that of a hard disk fluid with
effective area fraction $\eta=\frac{\pi \rho \sigma_{eff}^2}{4}$
and a mean-field correction which we simply model as
$\pi \rho^2 \int_{\sigma}^{\infty}{ \frac{U_a(R)}{k_BT} R
dR}$. For the free energy of a hard-disc fluid, analytical expressions are available
\cite{harddisks}. 
The free energy of the {\it solid} phase, on the other hand,  is
calculated as a lattice sum with the  
assumption of a two-dimensional triangular lattice structure. The lattice constant 
is directly related to the DNA number density $\rho$.

 Figure~\ref{figure_8} shows the resulting phase diagram with the
 coexisting DNA densities for  the wide range of $C$. At low $C$ there is  a
strong first-order gas-crystal 
phase transition whichs  coexistence region is widened for larger $C$ 
 due to the increasing attractions. Between the two threshold
 concentrations  $ C \approx 0.1$mM and $C\approx 65$mM there is enough 
attraction to stabilize a liquid phase of high DNA density. 
At higher Spe concentrations a second crystal,  with a considerable
larger lattice constant than that of the high-density solid
emerges. We call 
this novel phase a mesocrystal since its density is intermediate
between that 
of the fluid and the almost closed-packed solid.

Condensation (see the cross in
Figure~\ref{figure_8}) and subsequent redissolution (dot-dashed line in
Figure~\ref{figure_8}) of dilute DNA solution are  other implications
of phase diagram.  
As the spermine concentration increases for fixed  $\rho \sigma^2=0.002$, which 
 corresponds to a typical DNA concentration of 1mg/ml DNA, 
first the gas-liquid coexistence
line is encountered. This is the condensation transition and the system will split into a low
density gas and a high density 
liquid phase.  At much higher $C$ the
coexistence line is touched again 
and the system redissolves back into the dilute gas phase. The 
corresponding threshold
concentrations of the condensation and redissolution are in the range 
$C_c\approx 0.3$mM and $C_d\approx 165$mM and agree well with the
experimental observations \cite{raspaud1999,pelta1996}. 

\section{Discussions and conclusions}
One issue we want to discuss is the dependence of the phase diagram
on the DNA length $\ell$.
Since $\ell$ is a  prefactor of the effective potential,
it plays formally the role of an inverse system temperature.
We have explored the phase behavior for  smaller and larger
 DNA segment lengths
of $\ell=5P$ and $\ell=100P$ respectively. Firstly, the stability
of liquid pocket depends sensitively on $\ell$; it disappears
completely for small $\ell$, but extends towards larger $C$ for
larger $\ell$. Secondly, the fluid coexistence density
shifts to considerably higher values for smaller $\ell$.
Hence, condensation and redissolution is prohibited for small
DNA-segment lengths. This is in line  with the experiments of 
 Ref.~\cite{tripletDNAgoobes2002},
 where a threshold value of $\ell\approx 15P$ for the minimal
length $\ell$ required for condensation is reported.
For a triplex DNA (three stranded DNA molecule) the minimal
 length is reported to be about 2$P$
\cite{tripletDNAgoobes2002}. 
 The difference between the minimal lengths for duplex and
triplex DNA segments arises from the DNA charge density 
\cite{deserno2003,Deserno_JPC_2001}. The higher the linear charge density, the stronger the
overcharging. This in turn results in the precipitation of triplex DNA at
spermine concentrations, where duplex DNA does not aggregate
\cite{tripletDNAgoobes2002,longtriplex1999saminathan}. 

The DNA fragments considered here have a length less than
the persistence length of the DNA molecule which is around 500\AA. 
That is why one can safely adopt a rigid rod assumption for the DNA molecule 
and avoid the intramolecular fluctuations (off the long
DNA axis). This is also in line with the experiments [11-16] where 
the DNA redissolution measurements were done for DNA segments which
are smaller than the persistence length. Hence the particles involved
are very stiff rods. 

We speculate that the
height of the energetical barrier between the two minima in the interaction
potential $U(R)$ could regulate the DNA segment
lengths in the crystalline structures observed in experiments. 
Imagine that the  solution consists of a mixture of DNA segments
of different  lengths (but still smaller than the persistence length
500\AA). At lower spermine concentration, when the interaction 
potential has  a single
minimum,  all DNA segments will form a bundle
with an average inter-particle distance of the order of
$R=28$\AA. However,
 when the energetical barrier develops at intermediate 
spermine concentrations with a height around
$0.5k_BT-1k_BT$ per pinch length, only short DNA segment could
overcome this repulsion. Thus, short segments will fall into
first minimum and form a dense hexagonal 
structure, while longer DNA segments will be trapped in the second
minimum and form more swollen, fluid-like structure, apparently a cholesteric
phase.    

The other issue is the orientational entropy of DNA molecules 
and its influence on the interaction potential and free energy 
of DNA solutions. The evaluation of the contribution to the free energy from the 
orientational entropy of two nonparallel DNA rods (intermolecular
fluctuations) in MD simulations is a tremendous task. 
In fact,  the DNA-DNA effective potential must be averaged over 
all possible orientations for two DNA.
Unfortunately, an implementation of tilted DNA molecules will break
the system symmetry in simulations. 
On the other hand, simulation of
 20P long molecular segments for the system parameters 
 invoked here still is a challenge. 
For the bundle phase, due to the deep attraction (more than 40 $k_BT$) between
the DNA rods of 20P length, all other than parallel configurations of DNA 
molecules have little statistical weight. Thus, only a tiny 
correction to the energy of bundle is expected.

For the dilute gas phase the orientational entropy for DNA rod 
has an upper limit of about $2k_BT$.  At lower DNA densities, when the average
DNA-DNA distance is of the order or less than the DNA length, 
this entropy should be added to the free energy of liquid phase.
On the length-scale of free energy, where both the liquid perturbation
 and lattice sum produced curves have deep minima about dozen $k_BT$
 at intermediate DNA densities, such a upward shift of liquid free
 energy at smaller DNA densities will have only a slight effect on the
 phase diagram for 20P long DNA fragments. In other words, 
a corresponding upward shift of 
the free energy of dilute DNA system will not strongly affect the 
coexistence spermine concentration values deduced from the 
double-tangent procedure. Hence, a length of 20P for the  DNA segment is enough to
safely neglect orientational entropy effects in the phase 
diagram shown in Fig.10.

In conclusion, we have calculated the influence of tetravalent counterions 
on the effective interactions and the phase diagram
of columnar DNA assemblies by primitive-model-type computer simulations
and statistical theories. We find that a small concentration of tetravalent
counterions induces DNA condensation. The layering of the strongly coupled
tetravalent counterions on the DNA strands yields an oscillatory
effective interaction potential with a double-minimum structure at intermediate
counterion concentrations. This explains the redissolution transition
and triggers a novel stable mesosolid. Our threshold concentrations are 
in good agreement with experimental data.

\ack
This paper is dedicated to Lothar Sch\"afer on the occasion of his
sixtieth birthday. 
We acknowledge a partial support of this work by the European Networks
of Excellence 'SoftComp'.

\begin{figure}
    \epsfxsize=15cm 
    \epsfysize=15cm
 ~\hfill\epsfbox{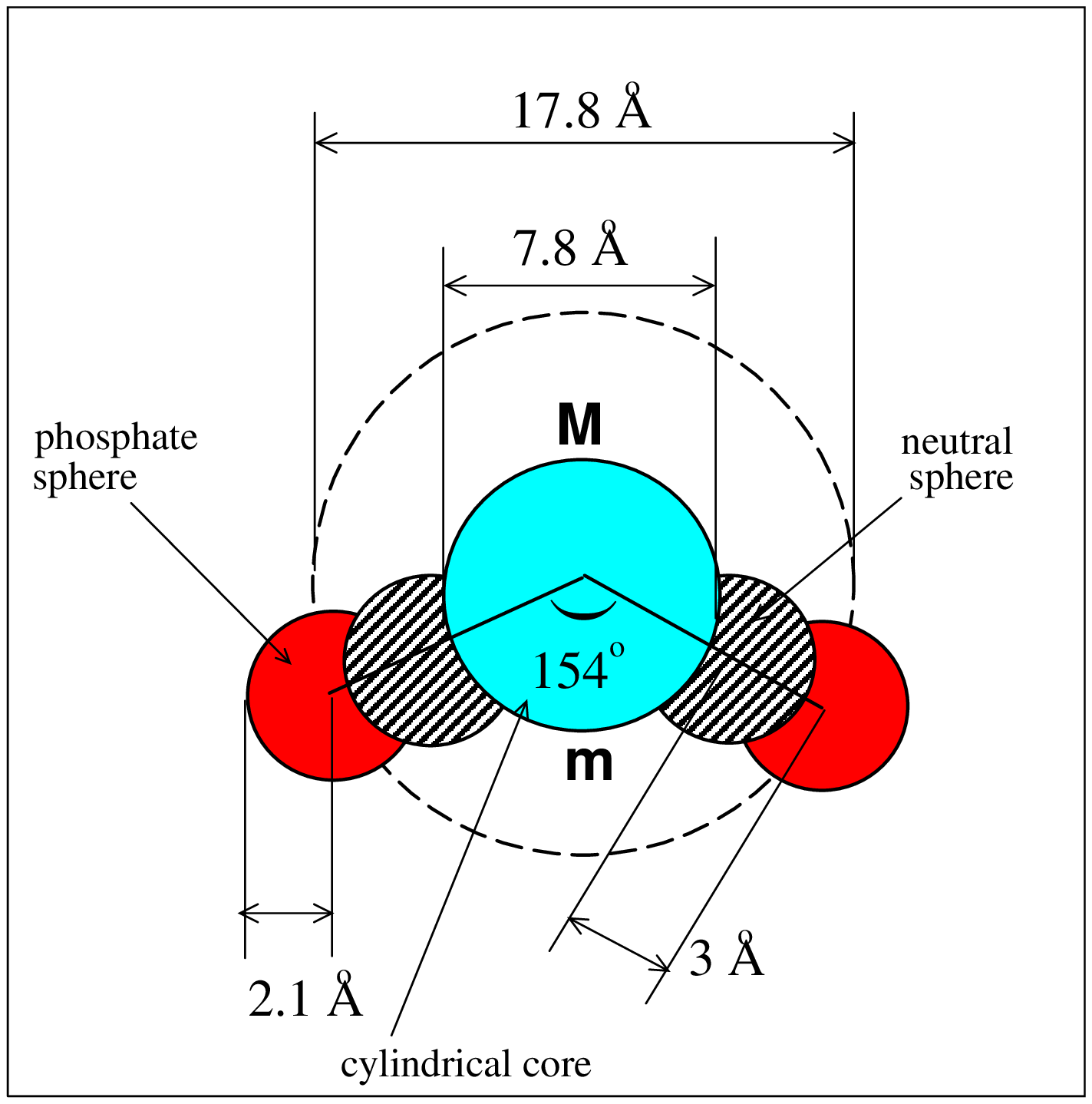} 
% \label{figure_1}
\end{figure}

\begin{figure}
   \epsfxsize=15cm 
   \epsfysize=15cm 
~\hfill\epsfbox{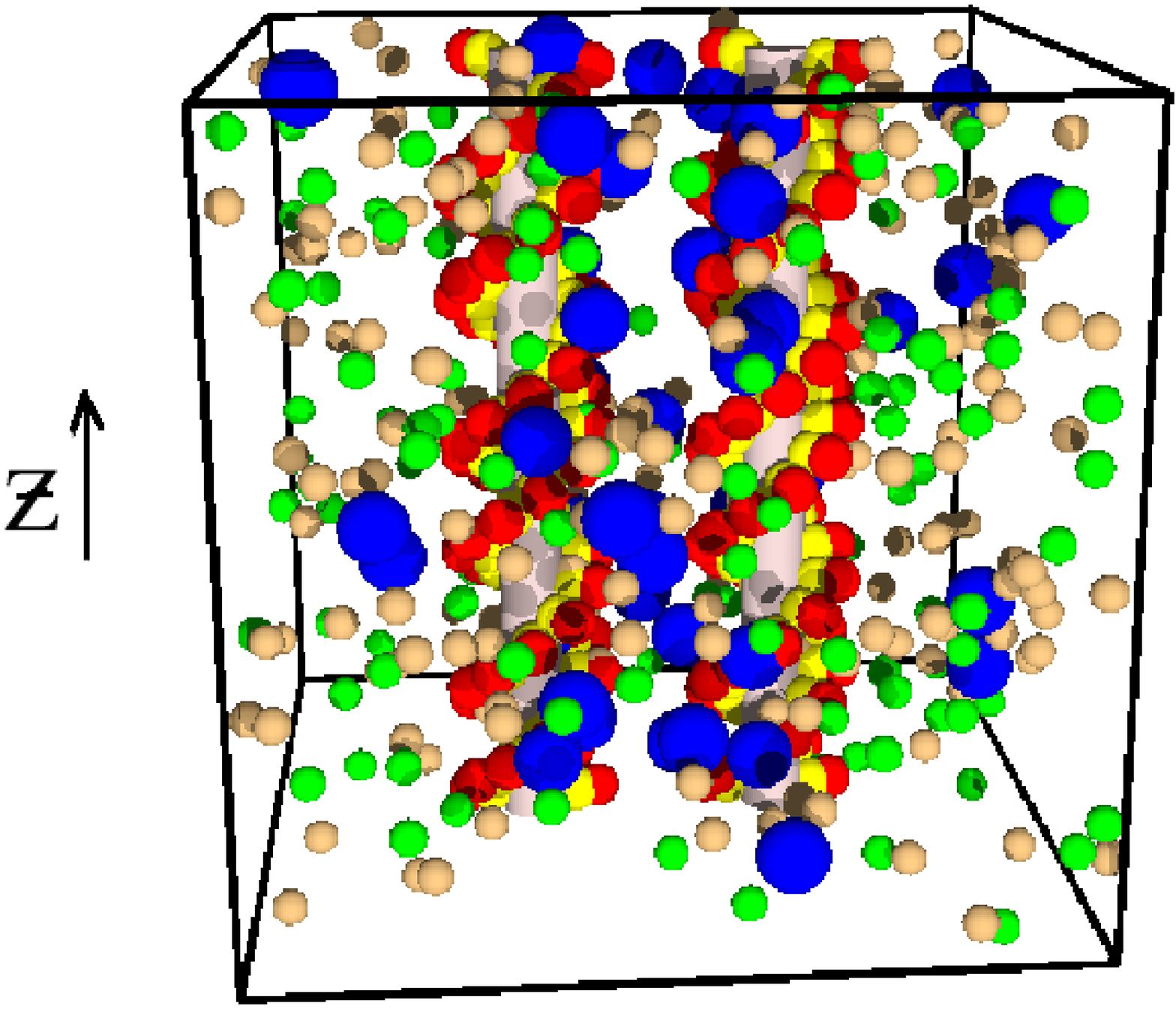}\hfill~   
% \label{figure_2}
\end{figure}

\begin{figure}
   \epsfxsize=15cm 
   \epsfysize=15cm 
~\hfill\epsfbox{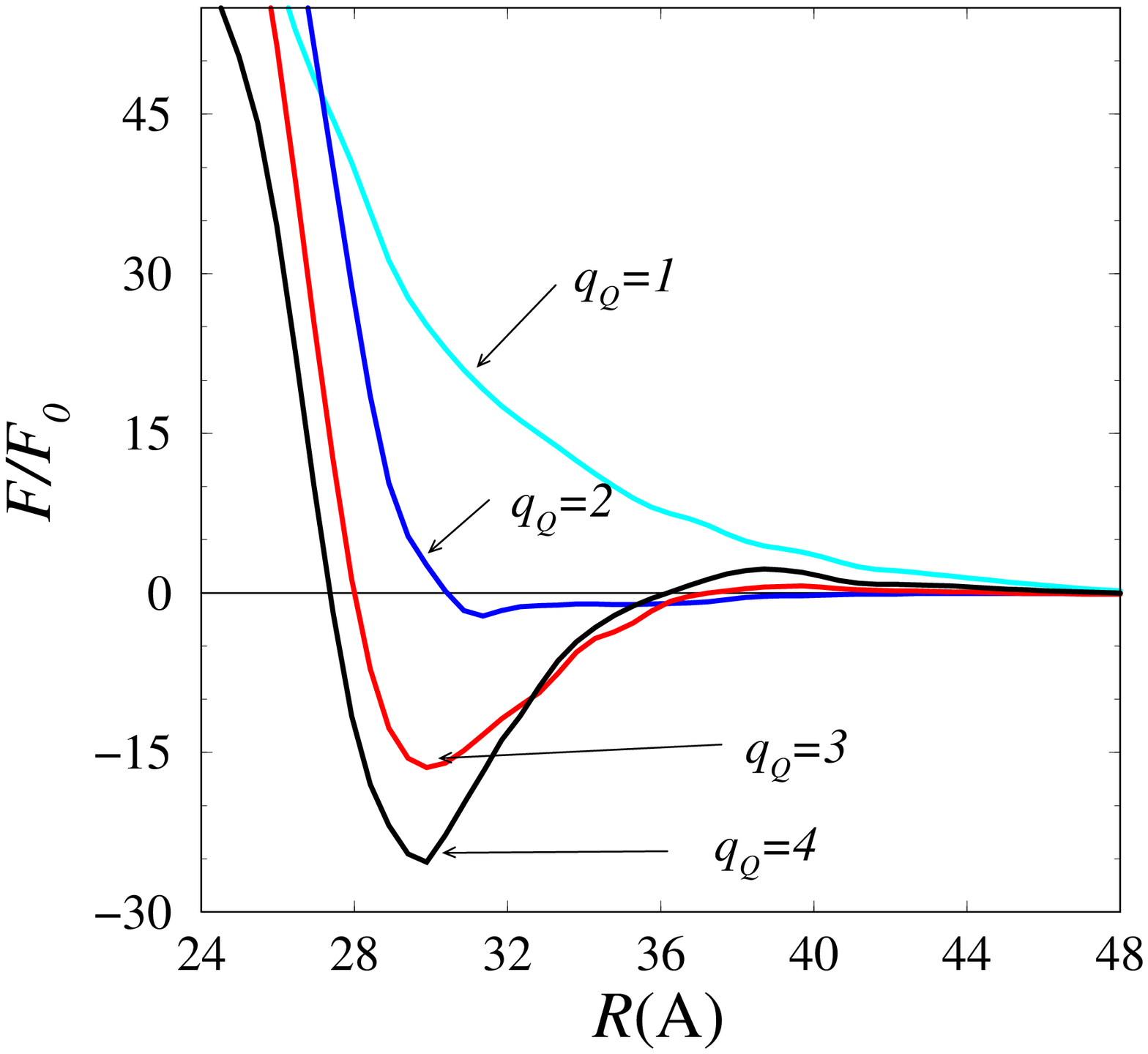}\hfill~
% \label{force_charge}
\end{figure}

\begin{figure}
   \epsfxsize=15cm 
   \epsfysize=15cm 
~\hfill\epsfbox{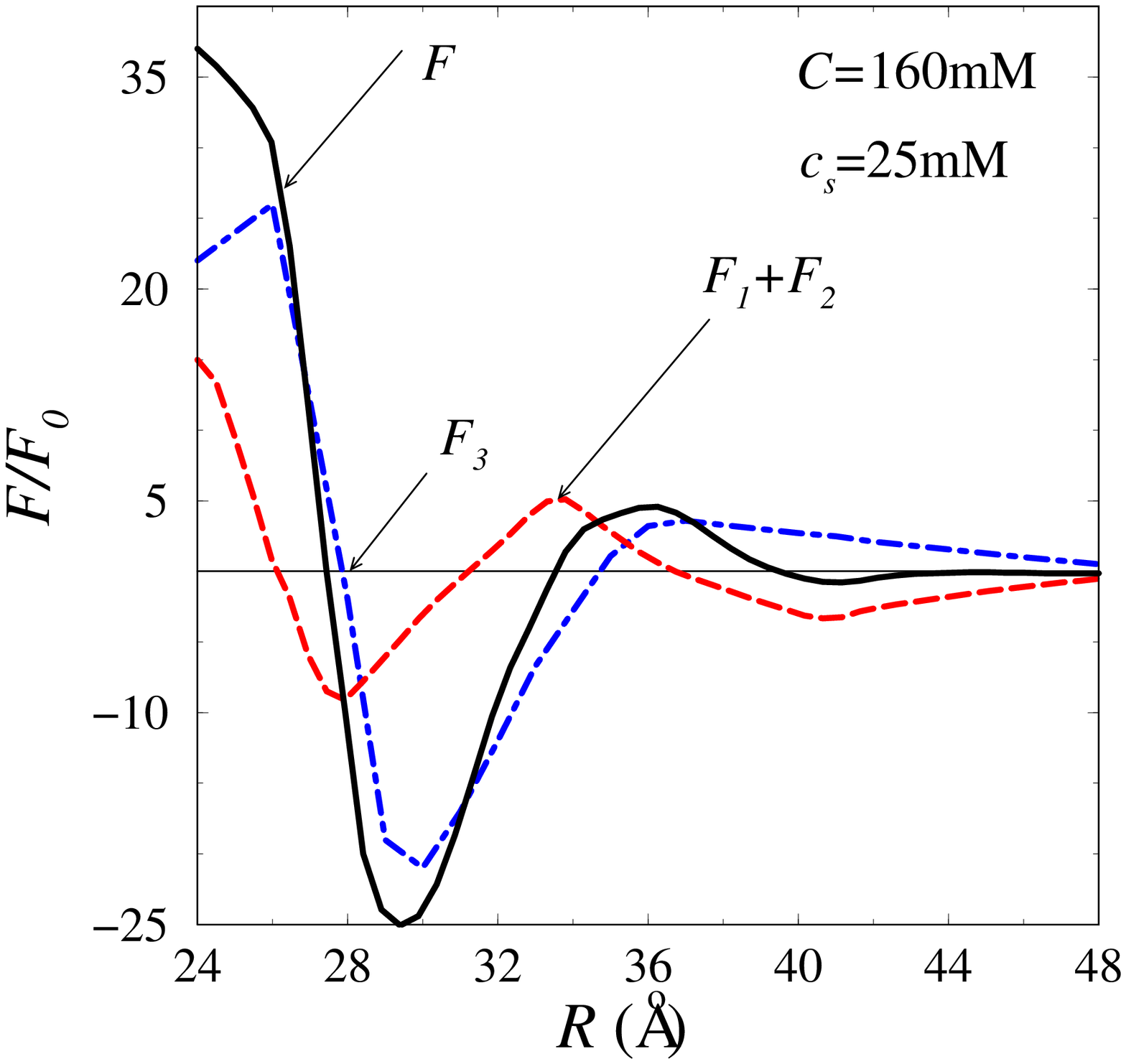}\hfill~
% \label{figure_4}
\end{figure}

\begin{figure}
   \epsfxsize=15cm 
   \epsfysize=15cm 
~\hfill\epsfbox{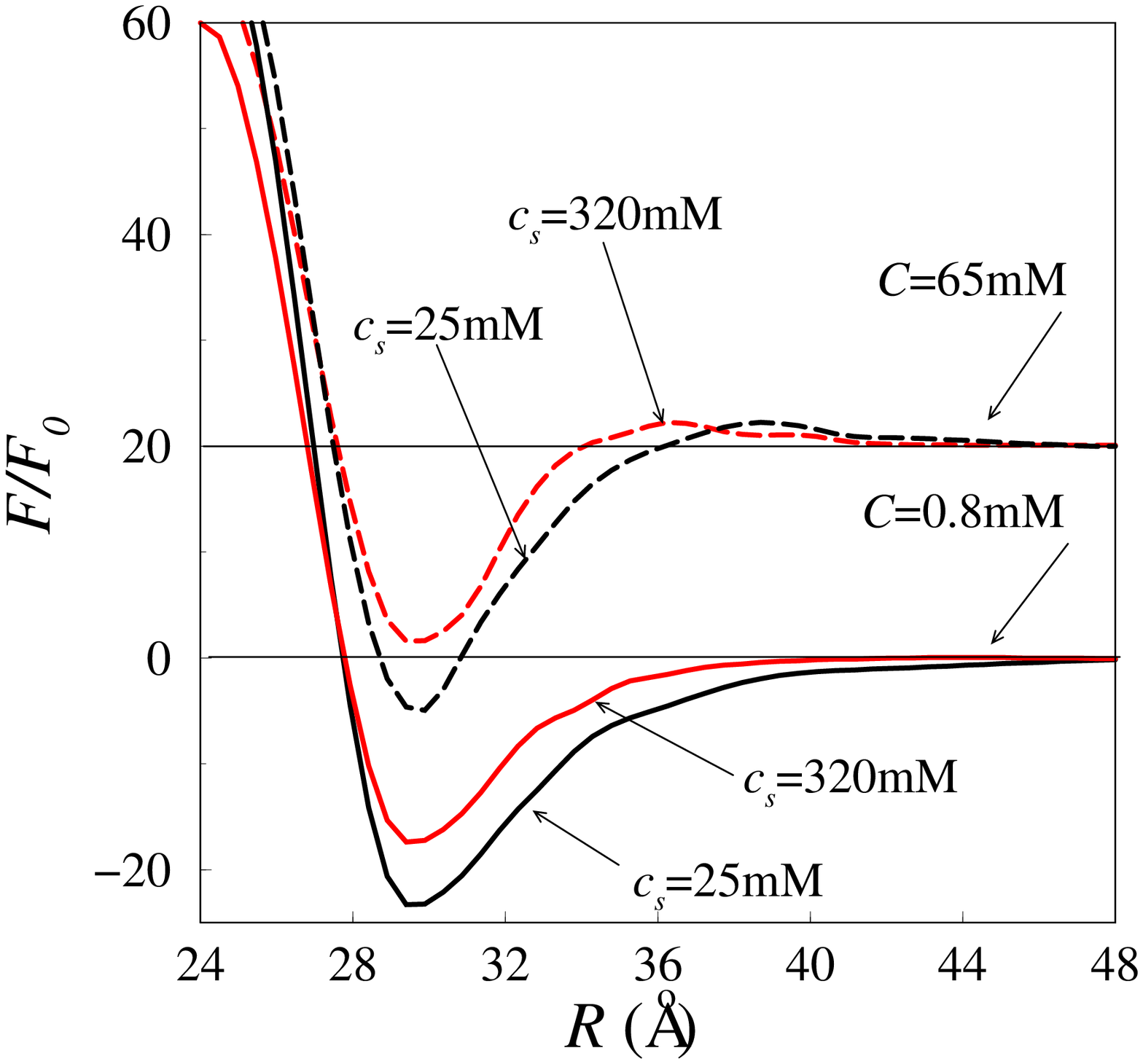}\hfill~
% \label{force_salt}
\end{figure}

\begin{figure}
   \epsfxsize=15cm 
   \epsfysize=15cm 
~\hfill\epsfbox{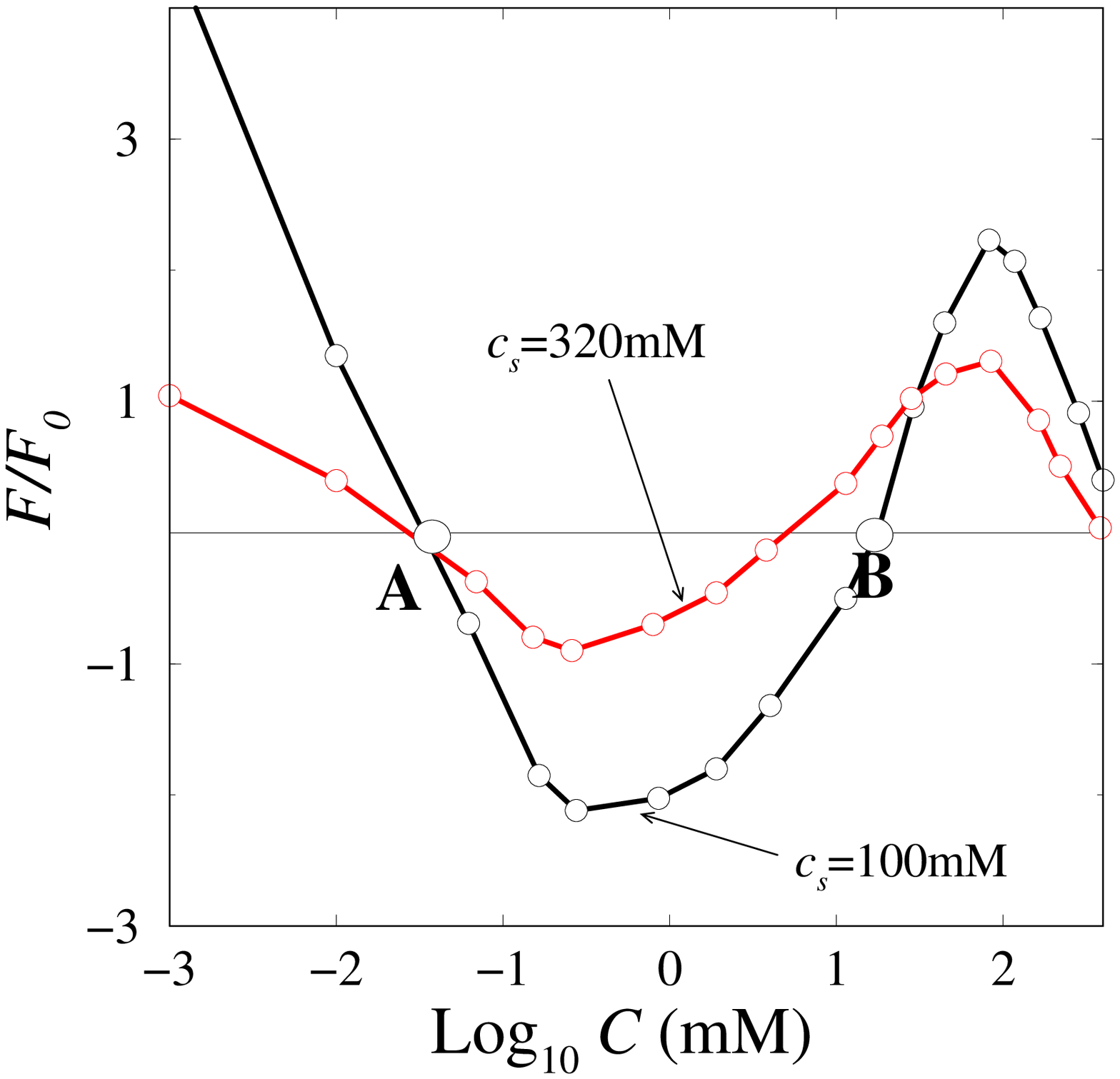}\hfill~
% \label{figure_5a}
\end{figure}

\begin{figure}
   \epsfxsize=15cm 
   \epsfysize=15cm 
~\hfill\epsfbox{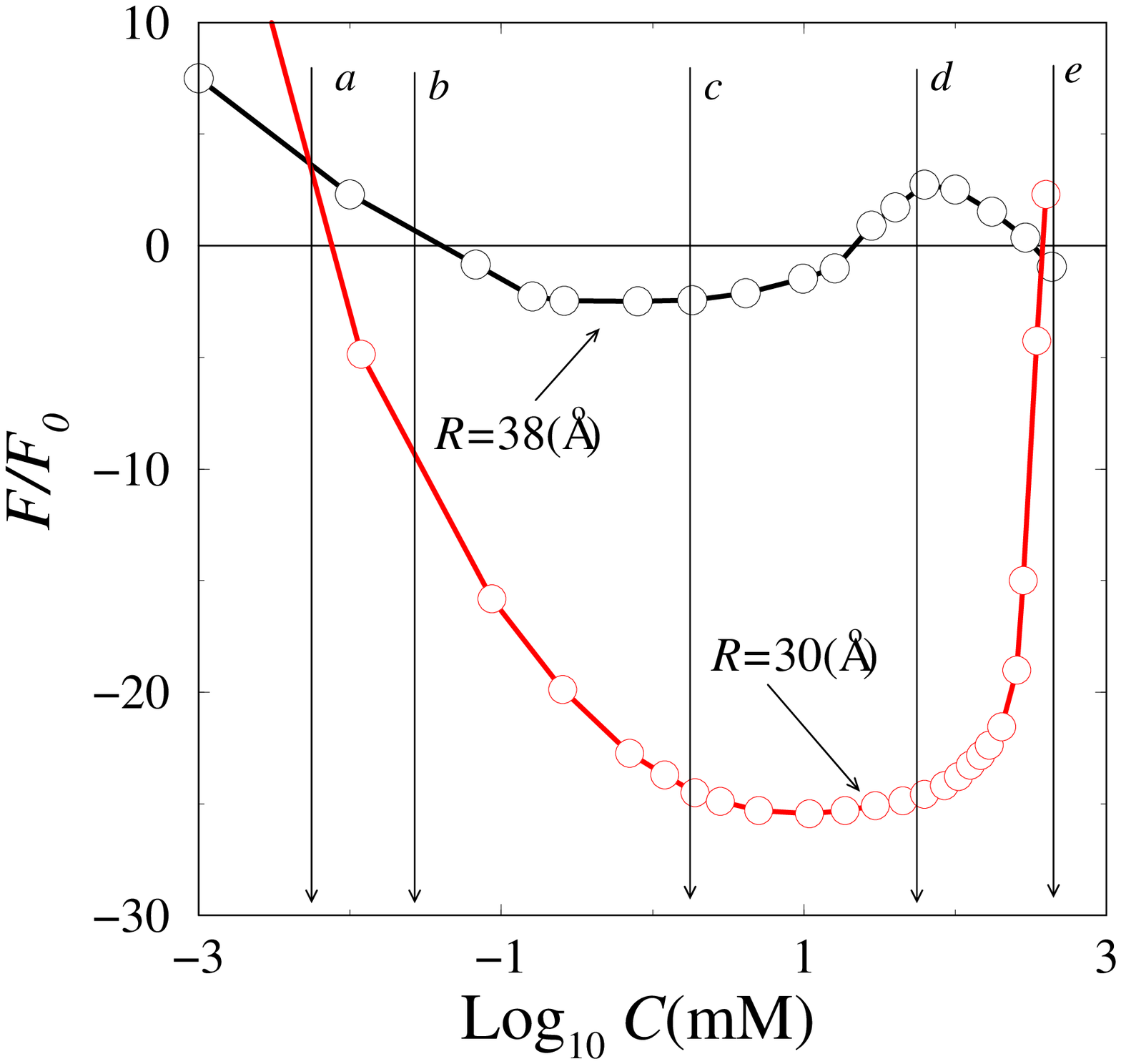}\hfill~
% \label{force_min_max}
\end{figure}

\begin{figure}
   \epsfxsize=15cm 
   \epsfysize=15cm 
~\hfill\epsfbox{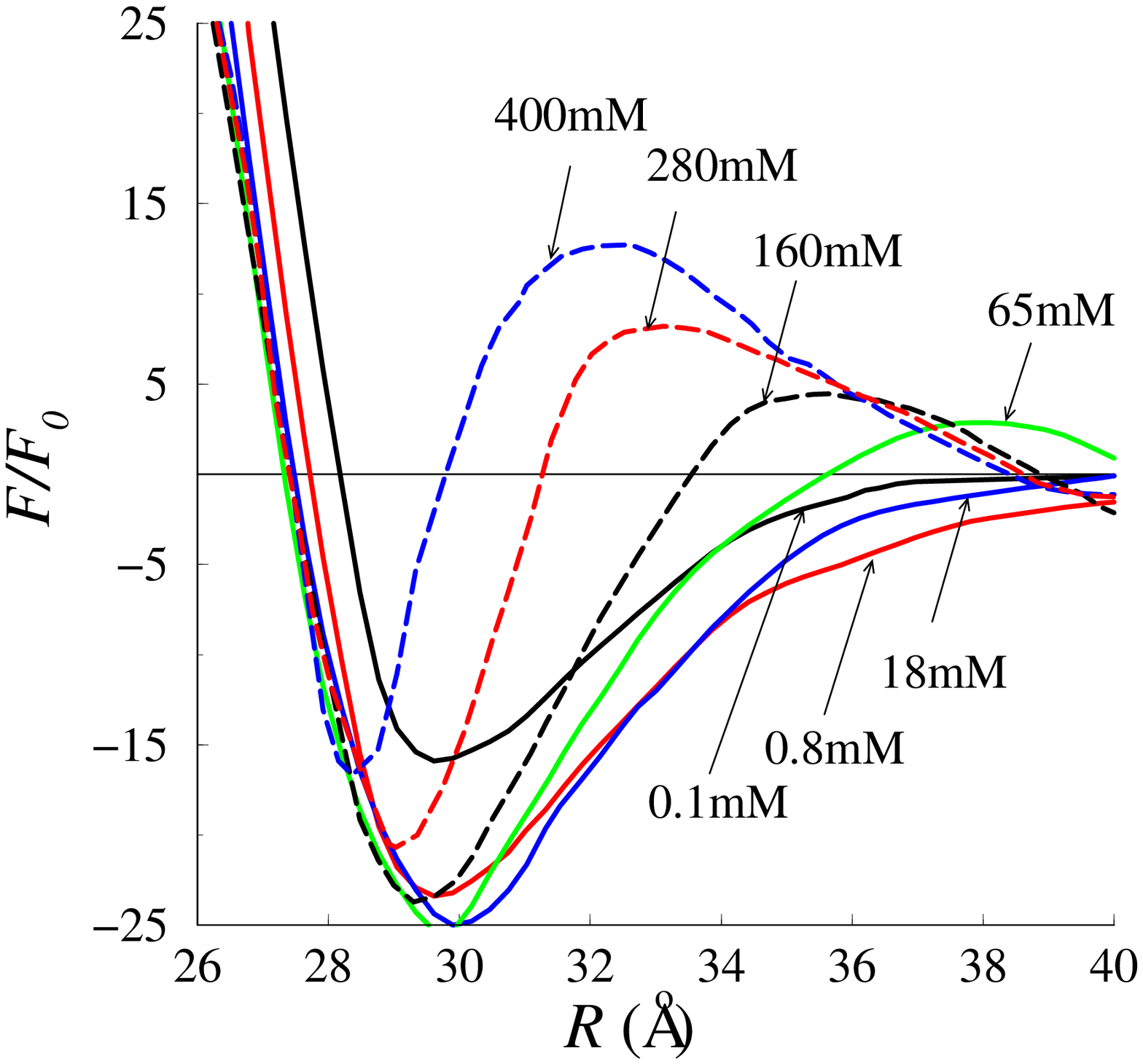}\hfill~
% \label{interctionforce}
\end{figure}

\begin{figure}
   \epsfxsize=15cm
   \epsfysize=15cm
~\hfill\epsfbox{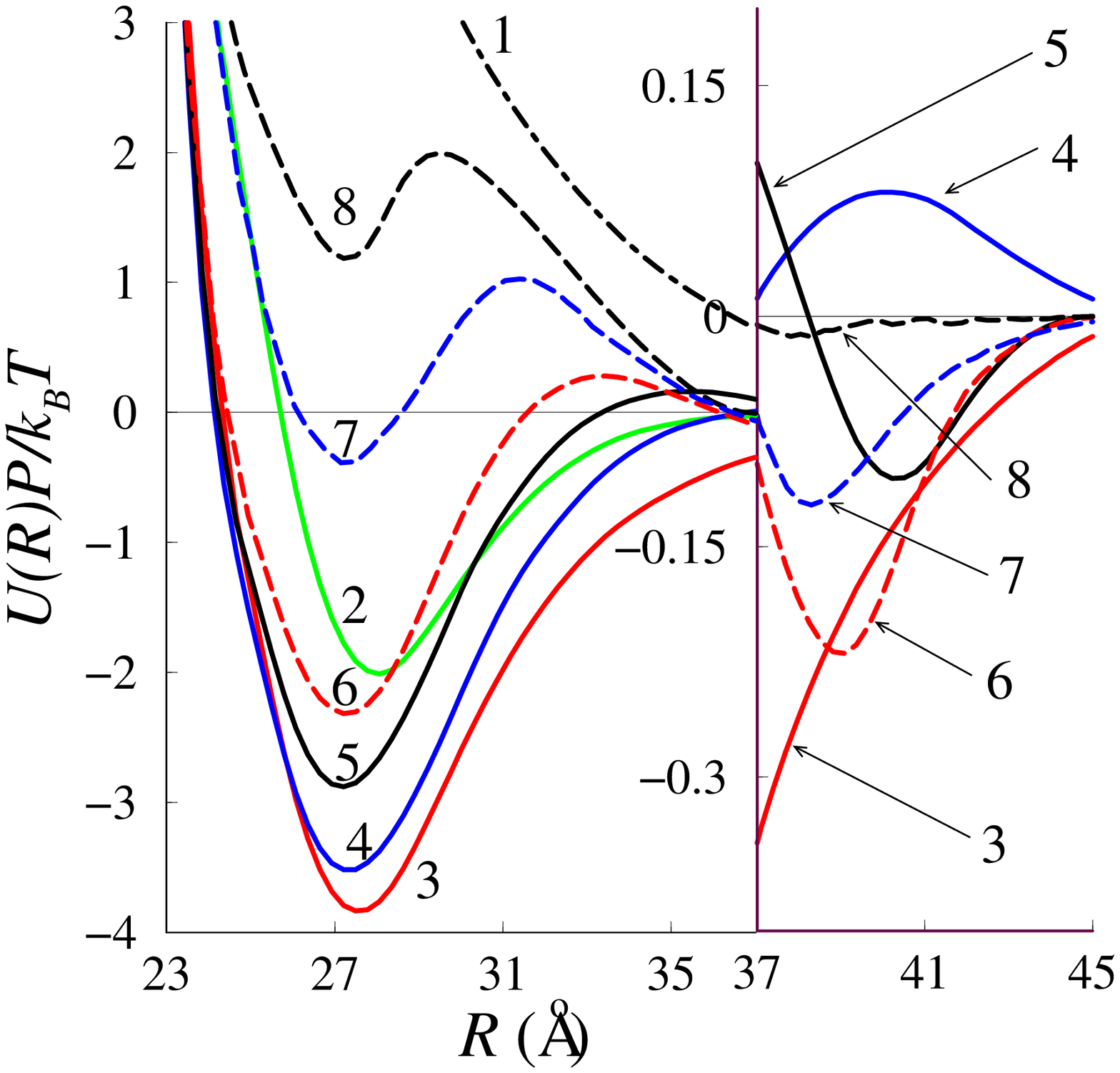}\hfill~
% \label{figure_5}
\end{figure}

\begin{figure}
   \epsfxsize=15cm 
   \epsfysize=15cm 
~\hfill\epsfbox{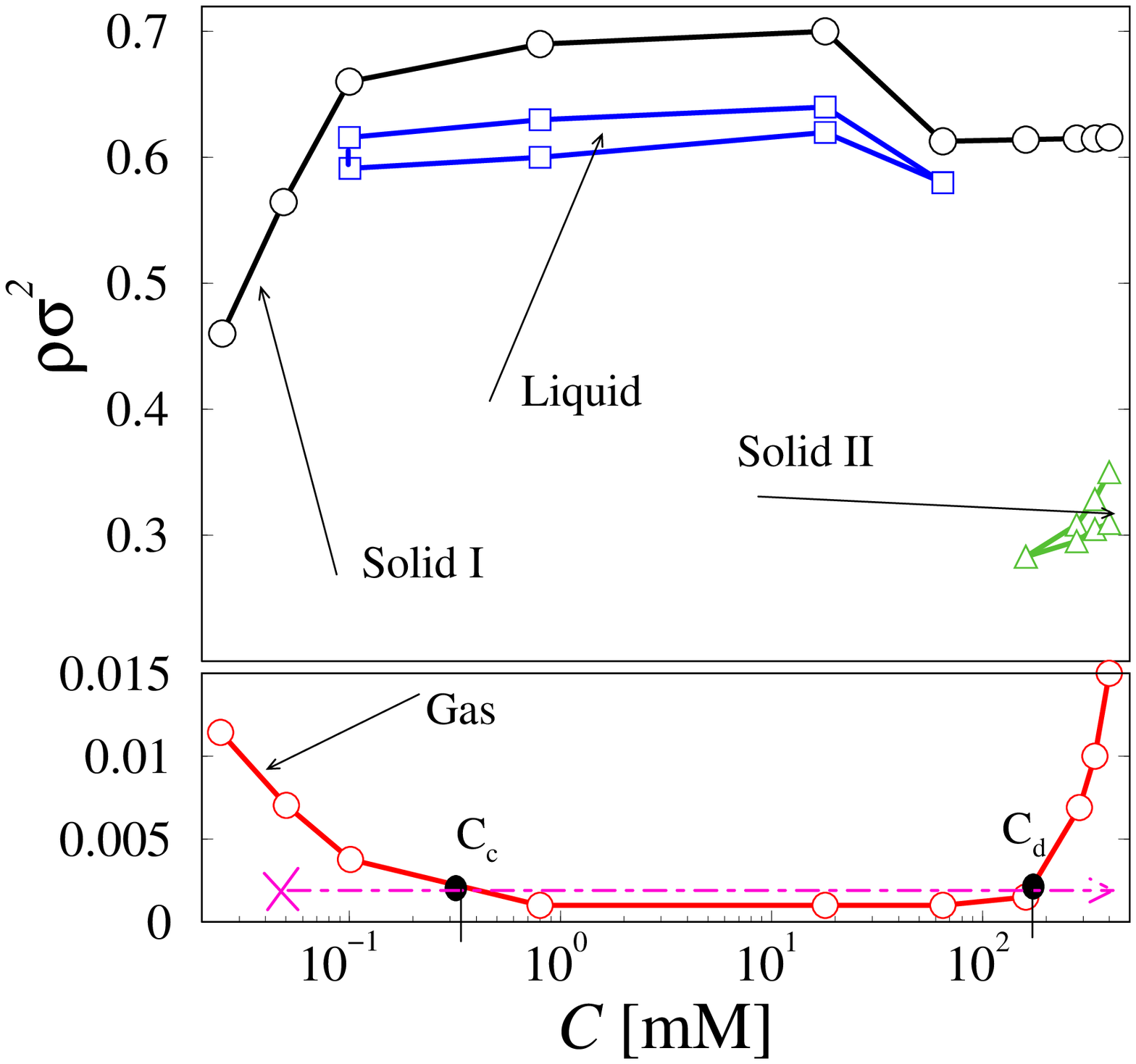}\hfill~
% \label{figure_8}
\end{figure}

\Figures
\begin{figure}
  \caption{(Color online) \label{figure_1} Cross section of  DNA in the $xy$ plane for
    the Montoro-Abascal model (MAM). Phosphate charges are
    shown as dark spheres. The DNA cylindrical core is colored in gray,
    the hatched areas correspond to neutral hard spheres. The
    inscribed letters "M" and "m" denote the major and minor grooves
    correspondingly. }
\end{figure}

\begin{figure}
 \caption{(Color online) \label{figure_2} Typical snapshot in the simulation cell.
The DNA molecules are shown as two parallel rods 
in the $\it z$ direction, over-wrapped by two
strings of light grey (neutral sphere in MAM (see text), colored yellow in online
figure) and grey (phosphate sphere in MAM,  red in online
figure) spheres. The tetravalent Spe ions are
   shown as big black (blue in online figure) spheres. Light grey
   (yellow in online figure) spheres represent
   coions, and dark grey (green in online figure) spheres are monovalent counterions.}
\end{figure}

\begin{figure}
  \caption{(Color online) \label{force_charge} DNA-DNA interaction force $F/F_0$ versus
   intermolecular separation distance $R$  for $c_s$=25 mM
   and $C$=65 mM. The charge of the big counterions is  indicated next to
  the  corresponding curves. $F_0=k_BT/P$, where $P=34$\AA.  }
\end{figure}

\begin{figure}
  \caption{(Color online) \label{figure_4} DNA-DNA interaction force $F/F_0$ versus
   intermolecular separation  distance $R$  for $c_s$=25 mM, 
   $C$=160 mM. 
 $F_0=k_BT/P$, where $P=34$\AA.
 The parameters
   corresponds to complete DNA overcharging. Note that there is a
   second maximum in the total force $F$ at
   about $R$=36 \AA \, and a second minimum at about $R$=41\AA. }
\end{figure}

\begin{figure}
  \caption{(Color online) \label{force_salt} DNA-DNA interaction force versus intermolecular
separation distance $R$ for $c_s$=25 mM
   (black lines in online figure)  
and $c_s$=320 mM (red lines in online figure) 
 and two different spermine
   concentrations: $C$=0.8mM (undercharged DNA, solid lines)  
    and  $C$=65mM (overcharged DNA, dashed lines).     
 $F_0=k_BT/P$, where $P=34$\AA. The curves that correspond to
   $C$=65mM are shifter upward for clarity. }
\end{figure}

\begin{figure}
  \caption{(Color online) \label{figure_5a} DNA-DNA interaction force versus spermine
   concentration $C$. The intermolecular separation distance is $R$=38\AA \,
 (roughly the position of the second
   maximum for $C$=65mM). In the region between the points A and B 
(for salt concentration $c_s$=100mM)
   first minimum of the interaction force
   develops. $F_0=k_BT/P$, where $P=34$\AA.
}
\end{figure}

\begin{figure}
  \caption{(Color online)  \label{force_min_max} 
DNA-DNA interaction force versus spermine
   concentration $C$ at fixed intermolecular separation distances $R$ for $c_s$=25mM. At
   $R=$30\AA \, ($R=38$\AA) the first (second) minimum emerges at
 intermediate spermine concentrations $C$ and
   then disappears at higher $C$. Arrows and labels  from $a$ to $e$
   next to them   are a guide to eyes to point the different spermine concentrations where 
the shape of the interaction force undergoes significant changes. For more 
details see the text. }
\end{figure}

\begin{figure}
  \caption{ (Color online) \label{interctionforce} DNA-DNA
interaction force versus intermolecular separation distance $R$ for $c_s$=25mM, 
   Different spermine concentrations are indicated next to
   corresponding curves: $C$=0.1mM, 0.8mM, 18mM,
   65mM, 160mM, 280mM, 400mM.  
 $F_0=k_BT/P$, where $P=34$\AA \, is the
   DNA pitch length. 
}
\end{figure}

\begin{figure}
  \caption{(Color online) \label{figure_5} DNA-DNA effective pair potential
versus intermolecular separation distance $R$ for $c_s$=25mM. The spermine
   concentrations are  
 $C$=0mM (1), 0.1mM (2), 0.8mM (3), 18mM (4), 65mM (5), 160mM
   (6), 280mM (7), 400mM (8). } 
\end{figure}

\begin{figure}
  \caption{ (Color online) \label{figure_8} Coexisting DNA densities as a function of  Spe
   concentrations $C$, for $c_s$=25mM. 
The stable phases found are gas-like, liquid and two triangular
   crystals with different 
lattices constants (solid I and solid II). All phase transition
   between these phases are of 
   first order. For the sake of better resolution at smaller DNA
   densities, the {\it y}-axis is expanded below $\rho \sigma^2<0.015$.}   
\end{figure}

\end{document}